%% file: permissions_policy.tex
\documentclass[sigconf, nonacm, 10pt]{acmart}
\input{preamble/macros}

\usepackage[english]{babel}
\usepackage{blindtext}
\usepackage{booktabs} % For formal tables
\usepackage{subfigure}
\usepackage{epsfig,endnotes}
\newif\ifecomment
\ecommenttrue
\ifecomment 
\newcounter{ECommentNum}
\stepcounter{ECommentNum}
\newcommand{\enote}[2]{({\color{blue}{#1:}\arabic{ECommentNum}\stepcounter{ECommentNum}: #2})}
\else
\newcommand{\enote}[2]{}
\fi

\begin{document}
\title{Reining in Mobile Web Performance with Document and Permission Policies}

%\titlenote{Produces the permission block, and copyright information}
%\subtitle{Extended Abstract}

%\author{Anonymous Authors}
\author{Byungjin Jun$^\dagger$ \quad Fabi\'an E.  Bustamante$^\dagger$  \quad  Ben Greenstein$^\star$ \quad Ian Clelland$^\star$}
\affiliation{$^\dagger$Northwestern University \qquad $^\star$Google}
%\author{Fabi\'an E.  Bustamante}
%\affiliation{Northwestern University}
%\author{Ben Greenstein}
%\affiliation{Google}
%\author{Ian Clelland}
%\affiliation{Google}

%\author{Byungjin Jun$^*$ \hspace{20pt} Fabi\'an E. Bustamante$^*$ \hspace{20pt} Ben Greenstein$^\dagger$ \hspace{20pt} Ian Clelland$^\dagger$ \vspace{5pt} \\ $^*$Northwestern University \hspace{20pt} $^\dagger$Google}

\renewcommand{\shortauthors}{Jun. et al.}

\begin{abstract}
%    \blindtext

The quality of experience with the mobile web remains poor, partially as a result of complex websites and design choices that worsen performance, particularly for users in suboptimal networks or devices. Prior proposed solutions have seen limited adoption due in part to the demand they place on developers and content providers, and the performing infrastructure needed to support them.
We argue that \textit{Document and Permissions Policies} -- an ongoing effort to enforce good practices on web design -- may offer the basis for a readily-available and easily-adoptable solution. In this paper, we evaluate the potential performance cost of violating well-understood policies and how common such violations are in today's web. Our analysis show, for example, that controlling for ~\textit{unsized-media} policy, something applicable to 70\% of the top-1million websites, can indeed reduce Cumulative Layout Shift metric.

\end{abstract}

%\begin{CCSXML}
%\end{CCSXML}

%\ccsdesc[500]{Computer systems organization~Embedded systems}

%\keywords{ACM proceedings, \LaTeX, text tagging}

\maketitle

\input{sections/introduction}

\input{sections/background}

\input{sections/policies}

\input{sections/methodology}

\input{sections/measurements}

\input{sections/related_work}

\input{sections/conclusion}

\bibliographystyle{ACM-Reference-Format}
\bibliography{references}

\end{document}

%% file: preamble/macros.tex
\newif\ifcomment
\commenttrue
\newif\ifacm
\acmtrue
\newif\ifcameraready
\camerareadyfalse
\newif\ifwatermark
\watermarkfalse

\ifwatermark
      \usepackage{draftwatermark}
      \SetWatermarkScale{.7}
      \SetWatermarkText{\shortstack{In Submission:\\Do Not Distribute}}
      \SetWatermarkColor[rgb]{1,0.88,0.88}
\fi

\ifcomment

    \newcounter{MVNumberOfComments}
    \stepcounter{MVNumberOfComments}
    \newcommand{\mvnote}[1]{\textcolor{blue}{\small \bf [MV\#\arabic{MVNumberOfComments}\stepcounter{MVNumberOfComments}: #1]}}
\else
    \newcommand\mvnote[1]{}
\fi

\newcommand{\parax}[1]{\noindent \textbf{#1:}}
\newcommand{\eg}{{e.g.,}\xspace}
\newcommand{\ie}{{\it i.e.,}\xspace}

%% file: sections/introduction.tex
\section{Introduction}
\label{sec:intro}

%General statement introducing the broad research area / context 

Quality of experience (QoE) with the mobile web remains suboptimal~\cite{danielan:mobilespeed}, with the majority of pages taking several seconds to load~\cite{httparchive:speed}, years after users have moved to mobile devices as their primary way to access the web~\cite{statcounter:press}. For users connected over challenged networks with somewhat older devices, a common case in many developing countries~\cite{gsma:state2020, ahmad:developing}, this is painfully obvious. % FEB would be good to have references for this

%An explanation of the specific problem (difficulty, obstacle, challenge) to be solved​

Part of the problem is that much of the web has been designed, implicitly or not, for users on good networks and devices or, at least, without considerations of performance implications~\cite{butkiewicz:complexity}. This has resulted in more complex websites~\cite{httparchive:weight}, with heavy web fonts, external resources, large images, and animation that, while perhaps visually appealing for high-end users can be frustrating to the rest.
This is exacerbated by a very permissive web that turns a blind eye to not following best practices for implementing performant websites, even if such violations drastically worsen the mobile experience.                                                                             

%A brief review of existing or standard solutions to this problem and their limitations​
%Relevant literature is not cover at length here, just enough to build your argument​

This situation combined with the potential impact of poor web performance on user engagement (and profit)~\cite{brutlag2009speed} have served as motivation for a range of industry and academic efforts (e.g.,~\cite{vaspol:vroom, netravali:watchtower, amazon:silk, google:amp-firstpost}). Despite their demonstrated potential benefits, most solutions have seen limited adoption partially due to the demand they place on developers and content providers, and the performing infrastructure needed to support them. 
For instance, six years after its first release, the adoption of AMP, an effective~\cite{bjun:amp} and popular solution, is still around 0.2\% of all the websites~\cite{w3techs:ampusage}.

%An outline of the proposed new solution​
% Feature policies could help 
%  - developer, website developer, controling their own pages according to user's network and device conditions
%  - developers controling a third party page/content
%  - users controling a website, in general, perhaps in reaction to changing network conditions/device capabilities 

% Examples of policies
%  - unsizes media
%  - synchronous scripts

We argue that \textit{Document and Permission Policies}~\cite{feature-policy:bidelman} may offer a readily available, easily deployable, and effective way to ``nudge'' developers towards better, mobile performant practices. Feature Policy, as they were originally named, is a specification that allows developers to control certain features and APIs on a browser~\cite{permissionspolicy}.  The features referred to by these policies range from camera access and location information to image sizing and script execution. 

While originally designed for the developers, both developers and users could use feature policies to identify when best practices for better mobile web performance, such as disabling the rendering of unsized images that may require a layout shift or excluding synchronous scripts that may block rendering, are not being followed. These policies may form the basis of solutions that enable developers to adjust their content according to users' network conditions or device capabilities and/or control the content made available by a third-party provider, and allow end-users to control website as their network conditions change.

As a first step, in this paper, we evaluate the potential performance benefit of enforcing the best practices and how common the violations of such policies are in today's web. We identify a number of performance-associated policies and rely on microbenchmarks to evaluate the performance benefits of applying them. In our test, Speed Index can be improved by 1.4 seconds by applying \textit{oversized-image} policy, while \textit{font-display-late-swap} can boost Largest Contentful Paint by 3 seconds\footnote{The best practices and the level of the improvement depend on devices and network constraints.}.
To estimate the potential impact of this approach, we look at how frequently these good-practice policies are violated among the top-1million most popular sites. About 40\% of Alexa top 100 pages and as many as 70\% of Alexa top-1million violate the \textit{unsized=media} policy and more than 65\% of top-1million pages include blocking scripts. Overall, top-1million pages violate 7.57 policies on average. These findings show the potential for new research on how to build mitigations for violations of best practices.

%% file: sections/background.tex
\section{Background}
\label{sec:bg}

The low QoE with the mobile web combined with its potential impact on user engagement (and profit)~\cite{brutlag2009speed} have served as motivation for a range of industry and academic efforts that, despite their demonstrated benefits, have seen limited adoption. We argue that \textit{Document and Permission Policies}~\cite{feature-policy:bidelman} may offer a readily-available and effective alternative. 

\subsection{Permissions and Document Policies}
\label{sec:policies}

Permissions and Document Policies are specifications that allow developers to control certain features and APIs on a browser~\cite{permissionspolicy,documentpolicy}. They were initially proposed as Feature Policy~\cite{feature-policy:bidelman} in 2016 and are a standard being implemented by several browsers including Chrome, Firefox, and Safari. 
	
A policy-controlled feature is an API or behavior which can be enabled or disabled in a document, or delegated to embedded documents, by referring to it in a permissions policy. A permissions policy for a document is a set of features and a list of allowed domains to which the features can be delegated.
By enabling or disabling a policy, developers can corroborate that their web application will behave as intended even if any third-party content moves against the developer’s intention, as the applied policy can report when a best practice is violated and make the browser enforce that the entire page follows the good practices.

There are about 35 features (5 others have been proposed) in Permission Policy\footnote{The full list of policies can be found in their webpage~\cite{permissionspolicy:features}} and they vary from simple features like \textit{camera} and \textit{geolocation} to performance-related ones such as \textit{execution-while-out-of-viewport}. 

Document Policy~\cite{documentpolicy}, unlike Permissions Policy which focuses on the permissions with binary choices (\ie yes or no), covers policies that are more related to performance aspects such as \textit{synchronous scripts} and \textit{oversized images} with possibly ranged parameters (\eg 2.0X or 3.0X). Currently, there are 13 features in Document Policy\footnote{The full list of policies can be found in the Chromium code~\cite{documentpolicy:features}}.

In order to apply policies on a webpage, two options are provided for developers; header policies and container policies.
A header policy is a list of features to control, delivered via an HTTP header with the page response. This declares the given policies for the entire document. On the other hand, a container policy applies to a browsing context container (\ie iframes) with the "allow" attribute. In addition to header policies and container policies, there is a JavaScript API as well that can observe and revise the allowance of each feature in the client-side code.

%% file: sections/policies.tex
\section{Repurposing Policies}
\label{sec:repurpose}

While originally designed for the developers, both developers and users could use Document and Permission policies to enforce best practices for better mobile web performance. Leveraging these policies as a ``scalpel'', developers could adjust their content according to users' network conditions or device capabilities and/or control the content made available by a third-party provider, while end users could control the overall website as their network conditions change. 

Unlike all-or-nothing approaches, policies would allow content providers or users to selectively apply each option based on their preferences, allowing them to trade aspects other than performance such as design and usability. 

Document and Permission policies demand little from developers or users, is very lightweight, and requires no effort to adopt as it is readily available in some browsers (\eg Chrome). In those browsers,  simply manipulating the header of the server response can change these feature policy configurations when the site is loaded on a browser.

\subsection{QoE-impacting Policies}
\label{subsec:targets}

Although there are nearly 50 Document and Permissions policies, it is clear that not all could have a significant impact on mobile web QoE.  For instance, while controlling access to user's camera and audio are undoubtedly important features, they likely have no effect on the Speed Index of a website. 

A careful review study of the different features, however, reveals a subset of features potentially useful in terms of QoE, including: \textit{oversized-images}, \textit{font-display-late-swap}, \textit{unsized-media}, \textit{layout-animations}, and \textit{sync-script}. 

\textit{Oversized-images} limits the maximum size images can be, under the given container (\eg img tag) and viewport size, as unoptimized images are one of the main sources of wasted bandwidth.  When an image violates this feature, it is replaced by a placeholder. Other, similar, image-related policies such as \textit{lossy-images-max-bpp}, \textit{lossless-images-max-bpp}, and \textit{lossless-images-strict-max-bpp}, limit maximum bits per pixel and show a placeholder for the violations, potentially having a similar QoE impact as \textit{ oversized-images}.

\textit{Unsized-media} applies to all media objects, images and videos, and set them to a default size (300px*150px) if they do not have an explicitly stated size. This feature should impact QoE since when media above the fold are unsized, it causes a layout shift that likely impacts user experience. 

Web fonts can also block loading a page. More commonly, however, a heavy font can be swapped later, once it has been loaded causing a re-rendering of the page and negatively impacting web QoE. \textit{font-display-late-swap} prevents this by disabling the later swap of the font if its loading time is long (\ie over 100ms).

\textit{Layout-animations} control animations that update layout since they consume significant CPU resources.
When it is enabled, such animations just show the initial and final states of the animation instead of the constant changes.

Finally, \textit{sync-script} blocks synchronous Javascript (\ie Javascript code without async or defer tags) when the feature is on. Depending on the placement in the page, synchronous scripts can block rendering events when it is on the critical path, and it causes a significant delay on displaying the contents.
Therefore,  we find and use blocking Javascript instead of sync-script in the original policy, as they have little performance impact if they do not block DOM rendering process.

%% file: sections/methodology.tex
\section{Policy Experimentation Setup}
\label{sec:methodology}

In the following paragraphs, we describe the experimental setup we have built to evaluate the potential impact of enabling different policies, on a variety of pages and, potentially, over a range of network conditions.

\subsection{A Chrome Extension}
\label{subsec:extension}

To experiment with the performance impact of different policies, we built a dedicated Chrome extension that can manage the application of policies to the current page on the client side.
This approach has a number of advantages: Chrome is the browser that most closely follows the evolution of policies; a client-side extension lets us test different policies on any web page we choose; it is relatively straightforward to build a prototype to run our analysis of a policy’s impact on web QoE; and this model makes it available to run a network emulation beyond a mobile setting, so to expand our study to other configurations (e.g., in-flight wifi~\cite{rula:ifc}).
%This option has a number of advantages. For starter, Chrome is the browser that most closely follows the evolution of  policies. Also, a client-side application like this let us test different policies on any web page we choose. Third, it is  relatively straightforward to build a prototype to run our analysis of a policy's impact on web QoE. Last, this model makes it easy to run a network emulation on the desktop device, so to expand our study to other configurations (e.g., in-flight wifi~\cite{rula:ifc}).

While the extension is activated, whenever a web page is requested from a server, the extension modifies the header of the incoming HTTP response to add the given list of policies before the response begins to be processed. The overhead imposed by the extension is negligible, since modifying the incoming header is all this extension does, and further, this overhead is constantly added to every web request even when no policy is enabled (thus factoring out in the comparative analysis). With the policy header in the incoming server response, the browser imposes the enabled policies on the loaded page. Note that this can break the resulting page depending on the nature of a policy (\S~\ref{subsec:targets}). While this may not be a desirable result for some users, the main purpose of this test is to learn the upper bound on the performance improvement expected from enforcing a policy, and second, users on challenged networks and/or poor devices may prefer to access the targeting information with lower delays if at the cost of some design features.

\subsection{Experimental setup}
\label{subsec:setup}

We carry on our measurements, emulating a mobile environment, on a MacBook Pro 2020 with a stable wired network. 
The automated testing is established with Puppeteer which launches the actual Chrome browser without caching.
Lighthouse -- a widely adopted web performance auditing tool -- takes over control to configure the mobile environment and collect auditing results. In our test, we not only set mobile user-agent, but emulate a mobile viewport (Motorola G4) as some policies are affected by its viewport size (\eg \textit{oversized-images}). We also throttle mobile network and CPU configurations using the options provided by Lighthouse. Specifically, we use the ``Slow 4G'' configuration\footnote{Regular 4G is pretty fast, so we targets the fastest setting among the "slow" network. The performance benefit would be larger in the poorer network. } that represents the bottom 25\% of 4G connections and top 25\% of 3G connections; the latency is 150ms, and upload/download throughput is 750Kbps/1.6Mbps respectively.
A comparison between Lighthouse and Netem~\cite{linuxfoundation:netem} throttling shows roughly similar results in many networking scenarios~\cite{debugbear:throttling}, with last-mile network emulation. 

%As Netem emulation similarly represents mobile network~\cite{bjun:amp}, we rely on Lighthouse in mobile network throttling in this paper.

%% file: sections/measurements.tex
%********************************************************************************
\section{Policy Potential}
\label{sec:performance}

In the following paragraphs, we present evaluation results from an analysis of the potential impact of policies on mobile performance. We use both a set of synthetic and real pages in our evaluation. We employ synthetic pages to characterize the potential impact of individual policies and then carry on a study of policy violations on the 1-million most popular pages to estimate the potential value of this approach in the wild. 

%****************************************
\subsection{Tested Pages}
\label{subsec:dataset}

The synthetic pages we employ have objects in the first mobile viewport, with each object included in violation of a particular policy. The images included are not optimized (600-700KB) to see the performance impact on webpages in the wild, which are more vulnerable to performance degradation.

We also use the Alexa top-ranked pages to evaluate how commonly pages in the wild violate best practices in policies. We take the top 1-million pages and cluster them in 5 groups: Top-100 pages, pages ranked between 100 and 1,000, 1,000 and 10,000, 10,000 and 100,000, and 100,000 to 1 million. From each bin, 100 accessible pages are randomly sampled for our testing. We would expect that lower-ranked pages would have a larger number of violations, on average, than those higher in the ranking, since popular page owners (\eg Amazon) likely have more resources to maintain their pages and enhance the performance.

\begin{figure}[h!]
  \centering
  \includegraphics[width=0.34\textwidth]{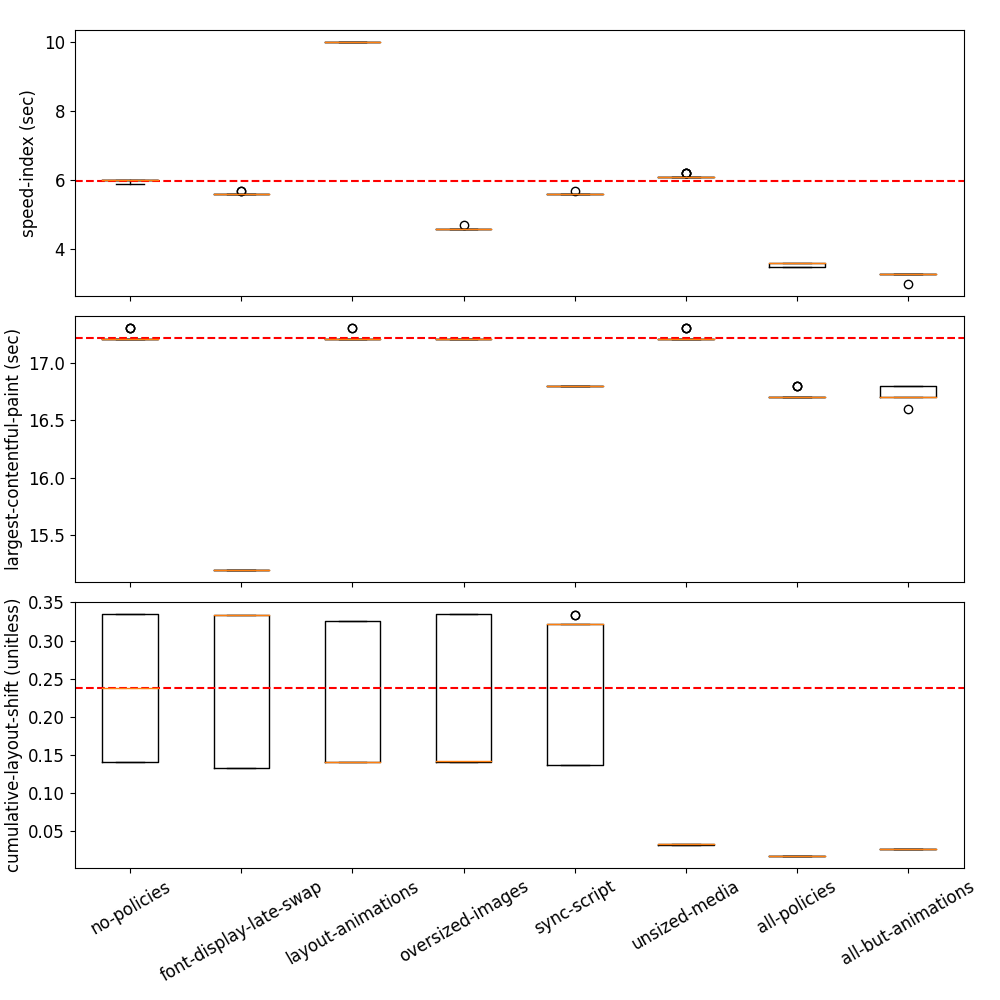}
  \caption{Performance impact applying each policy}
  \label{fig:performance}
\end{figure}

%****************************************
\subsection{Performance impact of policies}
\label{subsec:performance-impact}

\begin{figure}[h!]
  \centering
  \subfigure[Blocking JavaScript]{\psfig{figure=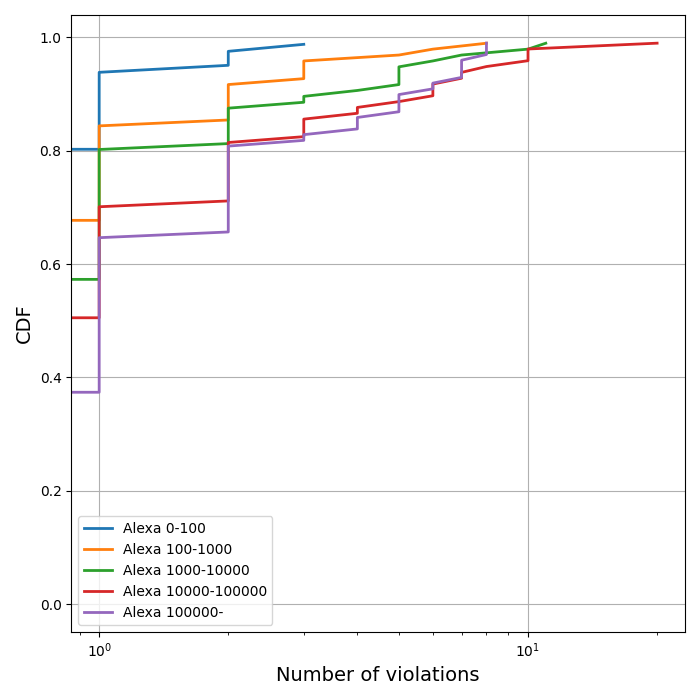, width=0.34\textwidth}\label{fig:js}}
  \subfigure[Unsized media]{\psfig{figure=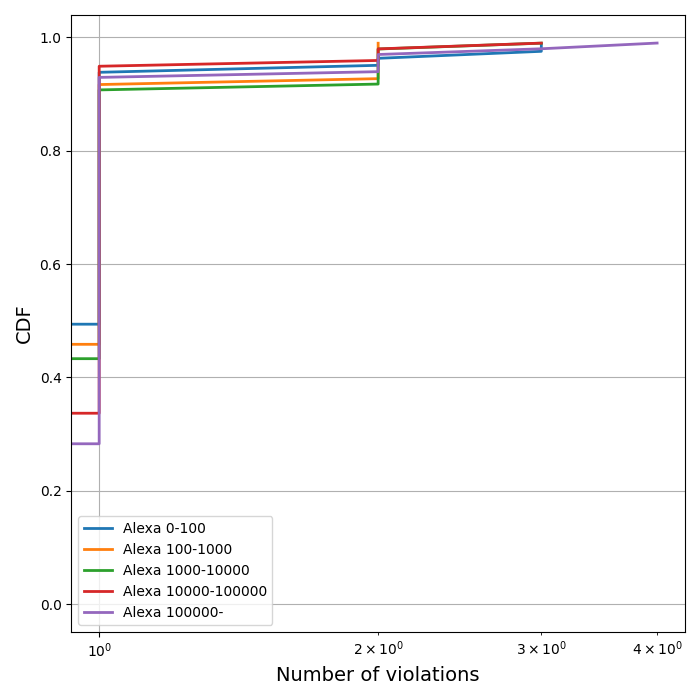,width=0.35\textwidth}\label{fig:unsized}}
  \caption{The example of the distribution of violations over different Alexa ranks in log scale}
  \label{fig:examples} 
\end{figure}

\begin{figure}[h!]
    \centering
    \includegraphics[width=0.34\textwidth]{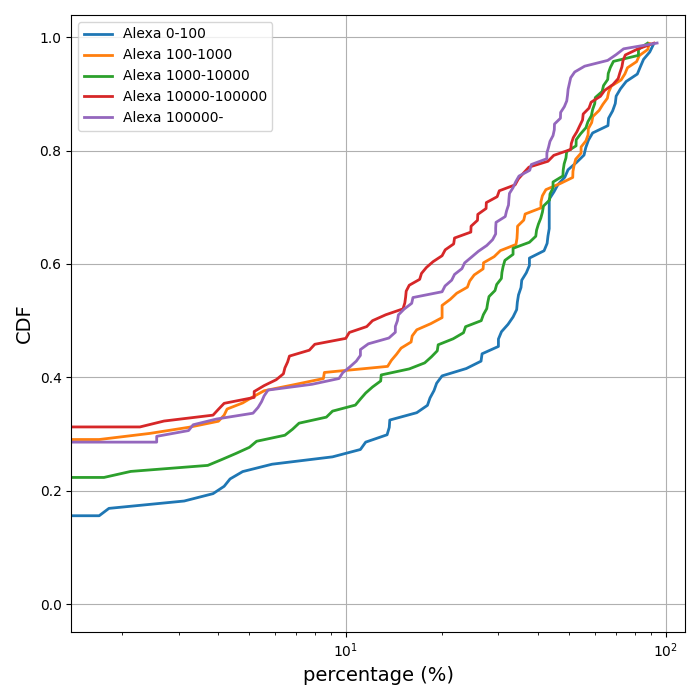}
    \caption{Speed Index improvement (\%) per page in log scale.}
    \label{fig:alexa-performance}
\end{figure}

We carry out control experiments on synthetic pages with specific policy violations and analyze their impact on Speed-Index (SI), Largest Contentful Paint (LCP), and Cumulative Layout Shift (CLS). Figure~\ref{fig:performance} presents these results based on 20 tests per selected feature. In addition to per-policy tests, the plots also include test results with all policies enabled and disabled as ground truth, and all-but-animations for further comparison. Overall, we find that while different policy violations impact different performance metrics differently, the performance impact of policy violations -- and thus the potential benefits of enforcing the associated best practices -- is consistent across metrics.

The top plot shows SI, which calculates when contents in a page are visibly populated with each policy applied. It is not surprising that large images degrade SI significantly, and synchronous scripts and large fonts can improve SI. The impact of \textit{layout-animation} policy, on the other hand, is surprisingly worse than \textit{no-policy}. We conjecture this is because SI calculation algorithm recognizes the drastic change (initial to final states) of the given animation after the policy is applied as a part of the rendering process. We reinforce this finding with the case of all policies, but \textit{layout-animation}, enabled. This shows even better performance than \textit{all-policies}. This is likely a matter of configuration of this policy, so its tuning will result in beneficial effects along with other performance-impacting features.

SI does not tell everything about the web performance, so we also look at LCP in the middle plot, which is one of the Web Vitals~\cite{webvitals}.
\textit{Font-display-late-swap} is interesting as it outperforms \textit{all-policies}.
We conjecture that this is due to the combination with other policies, particularly those associated with images, which are the key determinant of LCP.
In other words, image-related policies lead to some unknown negative interactions when working with \textit{font-display-late-swap}, thus \textit{all-policies} reveal worse performance than \textit{font-display-late-swap} only.
As expected, we find that the combination of \textit{font-display-late-swap} and \textit{sync-script} present the best result in an additional test.

\textit{Unsized-media}, which has little impact on other metrics, show greater improvement with CLS, simply because \textit{unsized-media} prevents layout shift by forcing all media to be sized.
Other test cases without unsized-media policy show two different values (around 0.15 and 0.33), as they can have a layout shift or two.
The one shift happens when the objects are loaded following the displaying order, and the page shifts twice when the next object is shaped earlier than the unsized media.

\subsection{How common are policy violations?}
\label{subsec:}

\begin{figure*}[h!]
	\centering
	\includegraphics[width=0.70\textwidth]{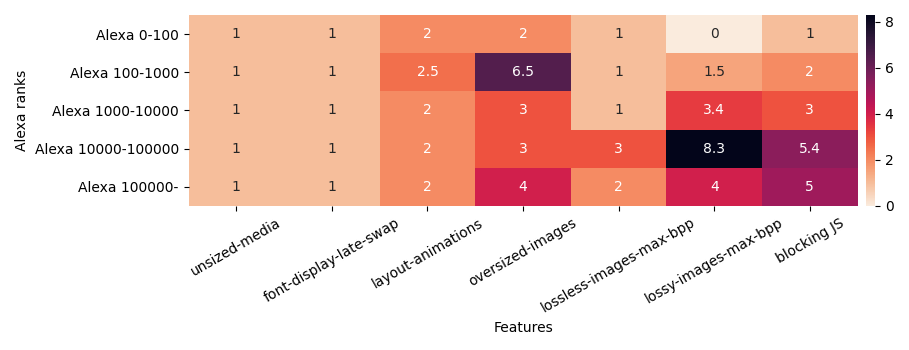}
	\caption{Common violations (90 percentile) per sampled bin of Alexa pages}
	\label{fig:heatmap}
\end{figure*}

Next, we collect the number of violations of best practices per page from five groups of sampled Alexa pages (\S~\ref{subsec:dataset}).
Figure~\ref{fig:heatmap} shows the number of violations of each policy per page in 90pct. We include \textit{lossless-images-max-bpp} and \textit{lossy-images-max-bpp}, which we didn't include in \S~\ref{subsec:performance-impact}, to see their frequency in the wild. Note that the number of image-associated violations is a conservative estimate, as we count only the first trigger violation (\eg an \textit{oversized image} violation, determined based on the container and viewport size, may hide an \textit{unsized image} violation). We find a very pervasive use of bad practices which, as expected, is even more common in lower-ranked pages.

We explore the distribution of violations of each policy across the Alexa ranks. Figure~\ref{fig:js} presents the results for blocking scripts across the Alexa top-1million. We find as many as 11 blocking script violations in the worst case. More than 60\% of the pages in Alexa top-100,000\textasciitilde 1,000,000 and over 20\% of Alexa top-100 pages have at least a blocking script. Figure~\ref{fig:unsized} shows the results for \textit{unsized-media}. The trend, in this case, is slightly different from blocking scripts - with fewer repeated violations per page but more pages with violations with 50\% in top 100 and 70\% in top 100,000\textasciitilde 1,000,000 pages showing a violation.

Figure~\ref{fig:alexa-performance} shows the performance improvement/cost, as measured by SI, for pages in the different subsets of Alexa top-1million pages of enforcing all the policies considered (ratio of SIs with every policy enabled over no policy).
Each Alexa subset shows overall enhancement with policies (70\%\textasciitilde 85\%) hinting at the potential impact of such an approach even for high-ranked pages.

%% file: sections/related_work.tex
\section{Related Work}
\label{sec:rw}

There is a significant body of prior work focused on improving mobile web performance, by reconstructing pages, resolving dependencies upstream or through client-side optimization. 

\parax{Reconstructing the pages} A popular approach is to rehabilitate the original page to optimize its performance.
Apple news~\cite{applenews}, Facebook Instant Articles~\cite{instantarticles} and Google Accelerated Mobile Project (AMP)~\cite{google:amp-firstpost} induce participating content providers to generate content fitting their standard (with good practices) to boost their pages as well as secure more users in their platform. AMP~\cite{google:amp-firstpost} provides content creators with a stripped-down and optimized version of standard web development tools so that they can benefit from multiple performance-improving techniques such as preventing blocking JS, caching, and pre-rendering. It has shown to yield significant performance benefits~\cite{bjun:amp}, although reconstructing existing pages may be demanding to minor content providers lacking resources.

\parax{RDR systems} Some efforts look to enhance mobile performance using Remote Dependency Resolution (RDR) scheme, which resolves the dependency graph of a page at proxy to reduce the burden on last-mile links. These efforts include popular record-and-replay tools~\cite{netravali:mahimahi}, the use of server's aid instead of proxy's~\cite{vaspol:vroom}, and recent works that enables RDR selectively~\cite{netravali:watchtower} and focus on third-party security~\cite{ronny:oblique}. The industry has proactively tried to provide their own solution for better performance~\cite{amazon:silk, opera:mini} using specialized web browsers to get RDR's help.
RDR solutions lead to significant performance enhancement, but its adoption in the wild is limited due to its requirement of powerful proxy.

\parax{Client-side optimization}
Several approaches look to expedite the loading process with the optimizations on the client-side, including caching and prefetching.
Caching applies in a diverse way, including the reduction of HTTP requests~\cite{mickens:silo}, mimicking mechanism or mobile Apps~\cite{shaghayegh:fawkes} and reusing identical computation~\cite{ayush:rethink}. Prefetching~\cite{google:amp-firstpost, venkat:prefetch} typically relies on predicting user behavior which can enhance speed significantly, when it succeeds, but it is not only challenging but privacy-concerning, and possibly leading to wasted energy and data usage~\cite{lenin:prefetch}.

We argue that the ongoing work on policies can form the basis of readily deployable, server, or client-side solution, that requires no infrastructure, supports a gradual adoption of best practices, and can yield immediate benefits on mobile web performance.

%% file: sections/conclusion.tex
\section{conclusion}
\label{sec:conclusion}

We argue that \textit{Document and Permissions Policies} may offer the basis for a readily-available, easily-adoptable way to improve mobile web performance. In this paper, we show the potential performance cost of violating the well-understood practices encoded in these policies, and how common such violations are in today's web. Building on these findings, we are exploring approaches that can leverage policies to adaptively improve the mobile web experience of all users.